\documentclass[english,11pt,aps,nofootinbib,amsfonts,amssymb,superscriptaddress]{revtex4} 
\usepackage{amsmath}
\usepackage{babel}
\usepackage{bm}
\usepackage{enumerate}

\begin{document}

\title{Propagation of light in area metric backgrounds}

\author{Raffaele Punzi}
\email{raffaele.punzi@desy.de}
\affiliation{Zentrum f\"ur Mathematische Physik und II. Institut f\"ur Theoretische Physik, Universit\"at Hamburg, Luruper Chaussee 149, 22761 Hamburg, Germany}
\author{Frederic P. Schuller}
\email{fps@aei.mpg.de}
\affiliation{Max Planck Institut f\"ur Gravitationsphysik, Albert Einstein Institut, Am M\"uhlenberg 1, 14467 Potsdam, Germany}
\author{Mattias N.\,R. Wohlfarth}
\email{mattias.wohlfarth@desy.de}
\affiliation{Zentrum f\"ur Mathematische Physik und II. Institut f\"ur Theoretische Physik, Universit\"at Hamburg, Luruper Chaussee 149, 22761 Hamburg, Germany}
\begin{abstract}
The propagation of light in area metric spacetimes, which naturally emerge as refined backgrounds in quantum electrodynamics and quantum gravity, is studied from first principles. In the geometric-optical limit, light rays are found to follow geodesics in a Finslerian geometry, with the Finsler norm being determined by the area metric tensor. Based on this result, and an understanding of the non-linear relation between ray vectors and wave covectors in such refined backgrounds, we study light deflection in spherically symmetric situations, and obtain experimental bounds on the non-metricity of spacetime in the solar system.
\end{abstract}
\maketitle

\numberwithin{paragraph}{section}

\section{Introduction}
Refinements of metric geometry naturally emerge as
effective backgrounds from quantum electrodynamics \cite{Drummond:1979pp}, quantum gravity \cite{Gambini:1998it,Krasnov:2007uu,Krasnov:2007ky,Krasnov:2007ei}, string theory \cite{Schuller:2005ru}, spin foam models \cite{Maran:2005iu}, classical electrodynamics \cite{Tamm,Obukhov:2002xa,Hehlbook,Hehl:2002hr,Rubilar:2007qm,Lammerzahl:2004ww,Itin:2004za,Hehl:2004zk,Kaiser:2004mz}, and scalar tensor theories of gravity \cite{Punzi:2008dv}. A very natural, and surprisingly fruitful way to view all of these generalized geometries (or at least salient aspects in some cases) is in terms of area metric manifolds \cite{Punzi:2006nx,Punzi:2006hy,Schuller:2007ix}, where an area metric is a smooth covariant tensor field $G$ of fourth rank that assigns a measure to tangent areas in a way similar to how a metric assigns a measure to tangent vectors. 

Due to this emergence of area metric geometry as an effective spacetime structure in a variety of contexts, it seems worthwhile to study area metrics as a fundamental structure in their own right, abstracting the discussion from the different technicalities of the fields from which the structure arises.
In the present paper, we derive the equation governing light paths in such refined backgrounds. This question of light propagation is clearly of the utmost importance for the interpretation of observational data in the context of any theory in which an area metric background emerges, since most of what we infer about the large-scale structure of spacetime, we infer from lensing or redshift data of the light that reaches us. We find that the geometric-optical limit of Maxwell electrodynamics on area metric backgrounds admits an effective description of the propagation of light in terms of geodesics in a Finsler geometry~\cite{Shenbook}. The relevant Finsler norm is induced from the area metric, and is closely related to the local null structure of the manifold in terms of the area metric's Fresnel tensor.
While this local null structure is well-known \cite{Hehlbook}, the identification of its role in the differential equation governing light paths requires the proof of some further non-trivial results, and is the key technical achievement of the present paper.  

 In the light of recent studies of Finsler geometries in connection to the quantization of deformed general relativity~\cite{Girelli:2006fw,Vacaru:2007nh}, quantum generalizations of the Poincar\'e algebra~\cite{Gibbons:2007iu}, and as phenomenological models \cite{Skakala:2008kf}, it is certainly interesting to note that an area metric structure of spacetime also leads a particular Finsler geometry when it comes to the description of the effective motion of light. 
 
The rigorous derivation of our new kinematical results from first principles is the central theme of section~\ref{sec_light}. We apply these results in section \ref{sec_lensingplanetary} to the study of light deflection in spherically symmetric area metric spacetimes. 

\section{Propagation of light}\label{sec_light}
The propagation of light rays can be studied from first principles as the geometric-optical limit of Maxwell theory, both in metric and area metric spacetimes. In the familiar metric geometry, the metric plays a threefold role by providing  (i) the background for matter field dynamics, (ii) the local light cone structure, and (iii) the length measure whose stationarity yields the the geodesic equation. In area metric spacetime, however, we will see that these three conceptually entirely different roles are indeed played by different structures, namely (i) the area metric itself, (ii) the totally symmetric fourth-rank Fresnel tensor associated with the area metric, and (iii) the Finsler norm associated with the Fresnel tensor. For area metrics that are induced from a metric, these three structures coincide, in accordance with the fact that area metric geometry is just a refinement of metric geometry. In this section, we develop the above insights in detail, starting from what is known about the local null structure of area metric spacetimes at a given point, over a careful analysis of the duality between light ray vectors and wave covectors, the differential equation describing light propagation, and finally the very important special case of birefringent backgrounds. The propagation equation for light rays is the central result of this section, since it will enable us to study light deflection in area metric spacetimes in section \ref{sec_lensingplanetary}.

\subsection{Area metrics and local null structure}\label{sec_localnull}
A number of facts about the local null structure of area metric manifolds are known~\cite{Tamm,Schuller:2007ix}; originally they have been derived in the context of pre-metric electrodynamics \cite{Obukhov:2002xa,Rubilar:2007qm}. We briefly collect and elaborate on these results, before proving a powerful new theorem on the duality of ray vectors and wave covectors in the following section \ref{sec_raywave}. 

Recall that an area metric manifold~$(M,G)$ is a smooth differential manifold $M$ equipped with a smooth covariant rank four tensor field $G$ with the symmetries $G_{abcd}=G_{cdab}$ and $G_{abcd}=-G_{bacd}$. Moreover, an area metric is required to be invertible in the sense that there is a smooth inverse $G^{abcd}$ so that $G^{abpq}G_{pqcd}=2(\delta^a_c\delta^b_d-\delta^a_d\delta^b_c)$. Geometric information about null vectors and null covectors at a given point of an area metric manifold is obtained by studying the geometrical-optical limit of the electrodynamical field equations~\cite{Punzi:2006nx}
\begin{equation}\label{Maxwell}
    dF = 0\,,\qquad  dH = 0\,,   
\end{equation}
where the constitutive relation between the field strength $F$ and the induction $H$ (which on metric manifolds is simply Hodge-duality) is now given by $H_{ab} = -\frac{1}{4}|\textrm{Det }G|^{1/6}\epsilon_{abmn}G^{mnpq}F_{pq}$ with $\epsilon$ being the totally antisymmetric tensor density determined by $\epsilon_{0123}=+1$. A wave covector field $k$ is a section of $T^*M$ satisfying the following two algebraic conditions:
\begin{equation}\label{wavecovector}
k\wedge F=0\,,\qquad k\wedge H=0\,.     
\end{equation}                          
These can be solved as follows: first we set $F=k\wedge q$ for some polarization covector $q\neq 0$; then we use the constitutive relation to find $G^{ijmn}k_ik_mq_n=0$. This equation in particular implies $G^{-1}(F,F)=0$, which makes $F$ a simple null two-form. Because of the symmetries of $G$ the polarization covector is determined up to a gauge transformation $q\rightarrow q+\lambda k$ and up to a rescaling $q\rightarrow\lambda q$. The covariant condition that non-trivial solutions $q$ are admitted is the Fresnel equation~\cite{Hehl:2002hr}
\begin{equation}\label{gencones}
\mathcal{G}^{ijkl} k_{i} k_{j} k_{k} k_{l} = 0\,,
\end{equation}
where the Fresnel tensor $\mathcal{G}$ only depends on the cyclic part $C^{abcd} = G^{abcd} - G^{[abcd]}$ of the inverse area metric~\cite{Punzi:2006nx}. Generically, equation (\ref{gencones}) represents a quartic surface. If the area metric spacetime takes the almost metric form $G^{abcd}=g^{ac}g^{bd}-g^{ad}g^{bd}+\phi |\textrm{det }g|^{-1/2}\epsilon^{abcd}$ for some axial scalar $\phi$ and $\epsilon^{0123}=-1$, then the Fresnel equation factorizes as $(g^{ab}k_ak_b)^2=0$. For $\phi=0$ this includes electrodynamics on a metric spacetime, and so gives the usual null cone structure for wave covectors.

On the surface of discontinuity determined by (an integrable) wave covector field~$k$, special tangent vectors are distinguished. These ray vectors are defined  through the algebraic conditions
\begin{equation}\label{lightray}
F(X,\cdot)=0\,,\qquad H(X,\cdot)=0\,,
\end{equation}
where $F$ and $H$ are the actual solution of (\ref{wavecovector}). Ray vectors provide in a background-independent way the direction along which the energy-momentum of the electromagnetic field flows from the point of view of a local observer; in the particular case of a metric background, the spatial part of a ray vector can be shown to coincide with the Poynting vector. On an area metric background, the physical energy-momentum vector is constructed from the effective energy-momentum tensor defined in \cite{Schuller:2007ix}, and this physical definition can be shown to coincide with the algebraic one above. The condition for the existence of a solution of the system (\ref{lightray}) also reduces to a Fresnel-like equation for the ray vectors $X$. Indeed, writing $F^{\sharp\,mn}=G^{mnpq}F_{pq}/2$, we may simply set $F^{\sharp} =X\wedge Q$ for some vector $Q\neq 0$ and it remains to solve $G_{ijmn}X^iX^mQ^n=0$. An immediate consequence of this equation is that $G(F^\sharp,F^\sharp)=0$; in the language of area geometry this shows $F^\sharp$ is a null area. Non-trivial solutions for $Q$ exist if the dual Fresnel equation
\begin{equation}\label{genconesvectors}
\mathcal{G}_{ijkl} X^i X^j X^k X^l=0
\end{equation}
is satisfied for the light ray vectors $X$, where the dual Fresnel tensor only depends on the cyclic part ${G^C_{abcd} = G_{abcd} - G_{[abcd]}}$ of the area metric $G$ and is defined as
\begin{equation}
\mathcal{G}_{abcd} = -\frac{1}{24} \left|\textrm{Det }G^C\right|^{-1/3} \epsilon^{ijkl} \epsilon^{mnpq} G^C_{ijm(a} G^C_{b|kn|c} G^C_{d)lpq}\,.
\end{equation}
Thus the dual Fresnel tensor defines the null structure for vectors on a generic area metric background. In case the constitutive relation arises from an almost metric area metric, the dual Fresnel equation factorizes as $(g_{ab}X^aX^b)^{2}=0$, providing again the standard null cones, as they would appear on a metric background. None of the preceding results on the local null structure of area metric manifolds is fundamentally new. In the context of pre-metric electrodynamics, this local structure has been studied in great detail in Ref. ~\cite{Hehlbook}. We now come to new results.

First, we present a very useful new result on the conformal equivalence of Fresnel tensors whose local null structures coincide. This generalizes the well-known statement on the conformal equivalence of two metrics that define the same null cones.

\vspace{6pt}\noindent{\bf Proposition.} \emph{Two Fresnel tensors $\mathcal G_1$ and $\mathcal G_2$ for which the null condition $\mathcal G_1(k,k,k,k)=0$ is equivalent to $\mathcal G_2(k,k,k,k)=0$ are conformally related, i.e., $\mathcal G_2=\alpha\mathcal G_1$ for some function $\alpha$. An identical statement holds for dual Fresnel tensors.}

\vspace{6pt}\noindent\emph{Proof.} First assume the `diagonal' values $\mathcal G_1(\lambda,\lambda,\lambda,\lambda)$ of $\mathcal G_1$ vanish for all covectors $\lambda$; then the polarization formula, which reconstructs the general symmetric tensor $\mathcal G_1$ from its diagonal values, see e.g.~\cite{Mujica:2006}, tells us that $\mathcal G_1$ vanishes. Hence also $\mathcal G_2$ vanishes, and both Fresnel tensors are conformally related. Now assume a covector $\lambda$ with $\mathcal G_1(\lambda,\lambda,\lambda,\lambda)\neq 0$ exists; then by hypothesis also $\mathcal G_2(\lambda,\lambda,\lambda,\lambda)\neq 0$. Choose a coframe $\{\theta^{\hat a}\}$ with $\theta^{\hat 0}=\lambda$. Then the homogeneous polynomials $\mathcal G_{1,2}(k,k,k,k)$ are of degree four in the variable $k_{\hat 0}$. Over the complex numbers we can decompose them into linear factors, so that
\begin{equation}
\mathcal G_{1,2}(k,k,k,k)=\alpha_{1,2} \prod_{i=1}^4 (k_{\hat 0}-\beta_{1,2}^i)\,
\end{equation}
where $\alpha_{1,2}\neq 0$ are functions on the manifold $M$ and $\beta_{1,2}^i$ also depend on $k_{\hat 1},\,k_{\hat 2}$ and $k_{\hat 3}$. Since all Fresnel null covectors of $\mathcal G_1$ and $\mathcal G_2$ coincide we must have $\beta_{1}^i=\beta_{2}^i$. Therefore the diagonal values of $\mathcal G_2$ and $\mathcal G_1$ are proportional with $\alpha_1/\alpha_2$. Using again the polarization formula, it follows that $\mathcal G_1$ is conformally related to $\mathcal G_2$.~$\square$

\vspace{6pt}In the following section we will derive the important new result that the Fresnel tensor and the dual Fresnel tensor may be used to map wave covectors to ray vectors, and vice versa, independent of the polarization.

\subsection{Duality map}\label{sec_raywave}
For abelian gauge theory, wave covectors $k$ and light ray vectors $X$ are defined by the algebraic conditions~(\ref{wavecovector}) and~(\ref{lightray}), respectively. In this section we will show under which conditions the area metric background provides a one-to-one map between the directions of light ray vectors and the corresponding wave covectors. It will turn out that it is indeed the Fresnel tensor, and the dual Fresnel tensor, which under specific circumstances provide this relation, independent of the polarization of the gauge field, see Theorem 1 below.

To see this, consider a wave surface with normal covector $k$ that satisfies the Fresnel equation (\ref{gencones}) everywhere. Since this equation was obtained as a solvability condition, this ensures the existence of fields $F$ and $H$ solving (\ref{wavecovector}) for the given $k$. However, there could be several solutions for gauge fields depending on different polarization covectors $q$. For convenience, we define a frame $\{e_{\hat a}\}$ and the dual co-frame $\{\theta^{\hat a}\}$ with $\theta^{\hat 0}=k$ and $\theta^{\hat 1}=q$. In this frame we have $F=\theta^{\hat{0}}\wedge\theta^{\hat{1}}$. Using $F_{ab}X^{b}=0$ from the definition of the light ray vector $X$ in (\ref{lightray}), we find
\begin{equation}
   X=X^{\hat{2}}e_{\hat{2}}+X^{\hat{3}}e_{\hat{3}}\label{X1}\,.   
\end{equation}
Since $G^{ijmn}k_ik_mq_n=0$, we have $G^{\hat{0}\hat{1}\hat{0}\hat{\alpha}}=0$, which allows us to calculate the only non-vanishing components of $H$ as $H_{\hat 0\hat\alpha}\sim (G^{\hat 2\hat 3\hat 0\hat 1},-G^{\hat 1\hat 3\hat 0\hat 1},G^{\hat 1\hat 2\hat 0\hat 1})$ up to factors that are irrelevant for our argument. Note that the case $G^{\hat{1}\hat{3}\hat{0}\hat{1}}=G^{\hat{1}\hat{2}\hat{0}\hat{1}}=0$ is  physically pathological, since then $H \sim F$, so any covector is a wave covector, and hence there is no null structure. We exclude this case in the remainder of this section. Using the results for the components of $H$ in $H_{ab}X^{b}=0$, we conclude that
\begin{equation}
G^{\hat{1}\hat{3}\hat{0}\hat{1}}X^{\hat{2}}=G^{\hat{1}\hat{2}\hat{0}\hat{1}}X^{\hat{3}}\label{X2}\,.
\end{equation}
It is now easy to see that up to rescaling the solution for $X$ is given by $X=G^{\hat{1}\hat{2}\hat{0}\hat{1}}e_{\hat{2}}+G^{\hat{1}\hat{3}\hat{0}\hat{1}}e_{\hat{3}}$. This relation can be rewritten, in a frame-independent manner~\cite{Rubilar:2007qm} as
\begin{equation}\label{Xvec}
 X=G^{-1}(\cdot,q,k,q)\,.
\end{equation}
This expression is gauge-independent: the substitution $q\rightarrow q+\lambda k$ has no effect because $G^{-1}(\cdot,k,q,k)=0$. A simple contraction shows once more that $k(X)=0$, so that the ray vector field $X$ is everywhere tangent to the wave surface; if one specifies the polarization vector on this surface, then the ray vectors are also unique, up to rescaling. At this point it would seem as if the map between vectors and covectors depended on polarization. That this is generically not the case will be shown in the following. For the special case of an almost metric background, equation~(\ref{Xvec}) yields $X^a=-(q_dq^d)k^{a}$. So, up to an irrelevant factor, $X=g^{-1}(\cdot,k)$. Note that, in this case, the vector $X$ does not depend on the polarization covector, in the sense that this dependence can be removed by an appropriate gauge choice. We will now study under what condition this familiar one-to-one correspondence between the wave covectors~$k$ and the ray vectors~$X$ holds for generic area metric backgrounds.

Writing the Fresnel equation again in the form $\textrm{det } G^{\hat 0\hat\alpha\hat 0\hat\beta}=0$ tells us that the $3\times 3$~matrix~$G^{\hat 0\hat\alpha\hat 0\hat\beta}$ does not have full rank. It is then clear that if and only if this matrix has rank two, the equation $G^{ijmn}k_ik_mq_n=0$ admits a unique solution for the polarization covector~$q$, at least up to a gauge redefinition $q\rightarrow \sigma q+\lambda k$ for $\sigma,\lambda\in{\mathbb R}$. In that case we expect $X$ to be determined by $k$ alone. Indeed, we are able to prove the following result on a one-to-one correspondence between the wave covectors and light ray vectors on general area metric spacetimes:

\vspace{6pt}\noindent\textbf{Theorem 1.} \emph{If $G^{\hat 0\hat\alpha\hat 0\hat\beta}$ has rank two, then the directions of the ray vectors~$X$ and wave covectors~$k$ are related by the Fresnel tensor as}
\begin{equation}
  X={\cal G}(k,k,k,\cdot)\,.
\end{equation}
\emph{Analogously, if the rank of $G_{\hat 0\hat\alpha\hat 0\hat\beta}$ is two then we obtain a relation via the dual Fresnel tensor:}
\begin{equation}
k={\cal G}(X,X,X,\cdot)\,.
\end{equation}

\vspace{6pt}\noindent\emph{Proof.} We prove the first relation; the frames are chosen as before. From the definition of the Fresnel tensor, we calculate the components of $\mathcal{G}(k,k,k,\cdot)=\mathcal{G}^{\hat 0\hat 0\hat 0 \hat a}e_{\hat a}$, and find
\begin{equation}
  \mathcal{G}(k,k,k,\cdot) \sim \left(G^{\hat 0\hat 2\hat 0\hat 2}G^{\hat 0\hat 3\hat 0\hat 3}-(G^{\hat 0\hat 2\hat 0\hat 3})^2\right)\!\left(G^{\hat 1\hat 2\hat 0\hat 1}e_{\hat 2}+G^{\hat 1\hat 3\hat 0\hat 1}e_{\hat 3}\right)  
\end{equation}
up to a nonvanishing rescaling factor. Comparing this to our previous result for $X$ in (\ref{Xvec}), it only remains to be shown that the first bracket is nonzero. Because of $G^{-1}(k,q,k,\cdot)=0$, the relevant submatrix of $G^{\hat 0\hat\alpha\hat 0\hat\beta}$ is the one for index values $\hat\alpha,\hat\beta=2,3$; the assumption that $G^{\hat 0\hat\alpha\hat 0\hat\beta}$ is of rank two corresponds to a non-vanishing determinant of the submatrix, but this determinant is precisely the first bracket above. Note here that $G^{\hat 0\hat\alpha\hat 0\hat\beta}$ being of rank two is equivalent to $\mathcal{G}(k,k,k,\cdot)\neq 0$, excluding the pathological situation where $G^{\hat{1}\hat{3}\hat{0}\hat{1}}=G^{\hat{1}\hat{2}\hat{0}\hat{1}}=0$. This proves the first relation; the dual relation can be proven in completely analogous fashion.~$\square$

\vspace{6pt}So far we have considered the local properties of wave covectors and light ray vectors on area metric spacetimes; in particular we have discussed the corresponding notions of null structures in any given tangent space, and their relation. The following section is devoted to the analysis of the differential properties of these objects, allowing us to generalize the concept of null geodesics to area metric spacetimes.

\subsection{Light trajectories}\label{sec_global}
We have seen that the Fresnel tensor, and the dual Fresnel tensor, determine the local null structure of the area metric spacetimes considered here. They also determine the relation between the wave covectors and the light rays. These results show that the Fresnel tensors play a role analogous to that played by a metric and its inverse in the special situation of a metric background. But the metric is also responsible for the propagation of light along null geodesics. This suggests asking whether the Fresnel tensor, or its dual, could also govern the propagation of light on general area metric backgrounds. Indeed, Theorem 2 below demonstrates that light rays propagate along those curves that stationarize a particular length functional defined in terms of the dual Fresnel tensor, which in four dimensions takes the form
\begin{equation}\label{actionH}
\int d\tau \,H(x,\dot x) = \int d\tau \,\mathcal{G}(\dot x,\dot x,\dot x,\dot x)\,.
\end{equation}
This result identifies Finsler geometry \cite{Shenbook} (for the Finsler norm $H^{1/4}$) as a useful tool in area metric geometry. 
Indeed, variation with respect to the curve produces the stationarity condition
\begin{equation}\label{Finslernull}
\partial_\tau\left(\mathcal{G}_{aqrs}\dot x^q\dot x^r\dot x^s\right)-\frac{1}{4}\partial_a\mathcal{G}_{pqrs}\dot x^p\dot x^q\dot x^r\dot x^s=0\,,
\end{equation}
and from the following theorem we learn that light rays are indeed described by solutions of (\ref{Finslernull}) for which
$H(x,\dot x)=0$ along the curve. Note that the action (\ref{actionH}) is not reparametrization-invariant, just like the corresponding action for null geodesics in Lorentzian geometry, since a reparametrization of the tangent vector~$\dot x$ adds a term of the form $\lambda\mathcal{G}_{aqrs}\dot x^q\dot x^r\dot x^s$ to equation (\ref{Finslernull}), where $\lambda$ is an arbitrary function. But since the null condition does not depend on $\lambda$, any choice of parametrization provides Finsler null geodesics. For generic area metric spacetimes, the familiar result from metric geometry, namely that light rays are described by null geodesics, generalizes to:

\vspace{6pt}\noindent\textbf{Theorem 2.} \emph{The integral curves of the light ray vector field $X$ are null geodesics with respect to the measure defined by $H(x,\dot x)$. This does not hold for points where the one-to-one correspondence between wave covector and light ray breaks down.}

\vspace{6pt}\noindent\emph{Proof.} The wave covector $k$ is normal to the wave surface, which we can imagine as the level set $\Phi(x)=0$ of some function. Then $k=d\Phi$, which implies $k_{[l}\partial_{m} k_{n]}=0$; due to the appearance of the tangential derivative $k_{[l}\partial_{m]}$ along the wave surface this expression is well-defined along a single level set of $\Phi$. By contraction with the light ray vector $X^l$, and using the fact that $k(X)=0$, we obtain
\begin{equation}\label{Proof1}
0=k_{[m}\left(X^l \partial_{n]}k_l-X^l\partial_{|l|}k_{n]}\right).
\end{equation}
Consider now points along the integral curve of $X$ for which ${\cal G}(X,X,X,\cdot)\neq
0$; as discussed in the proof of Theorem 1, on the correspondence between wave covectors and light rays, we then have $k={\cal G}(X,X,X,\cdot)$. Substitution into the equation above yields
\begin{equation}\label{Proof3}
0 =  k_{[m}\partial_{n]}{\cal
G}_{labc}X^a X^b X^c X^l + 3{\cal G}_{labc}X^a X^b X^c
k_{[m}\partial_{n]}X^l-k_{[m}X^l\partial_{|l|}{\cal
G}_{n]abc}X^a X^b X^c\,.
\end{equation}
We now use the fact that ${\cal G}(X,X,X,X)=0$ on the whole wave surface; hence its tangential derivative along this surface must vanish, so that
\begin{equation}\label{Proof2}
   0=k_{[m}\partial_{n]}{\cal G}_{labc}X^a X^b X^c
X^l+4{\cal G}_{labc}X^a X^b X^c k_{[m}\partial_{n]}X^l\,.   
\end{equation}
This relation allows us to replace the second term in (\ref{Proof3}) to obtain
\begin{equation}
 0 = k_{[m}\Big(\frac{1}{4}\partial_{n]}{\cal G}_{labc}X^a
X^b X^c X^l-X^l\partial_{|l|}{\cal G}_{n]abc}X^a X^b
X^c\Big)\label{Proof4}\,.
\end{equation}
The quantity in brackets is, up to a sign, the left-hand side of the equation of motion~(\ref{Finslernull}) for a Finsler null geodesic. Since we know this expression to be well-defined on the wave surface, it must have the form $-\lambda k_n$ for some function $\lambda$. This is precisely the term arising from some arbitrary reparametrization of the Finsler null geodesic, as discussed before. This completes the proof.~$\square$

\vspace{6pt}In the above theorem, we excluded points where ${\cal G}(X,X,X,\cdot)=0$. At those points,  equation~(\ref{Finslernull}) for Finsler null geodesics does not admit a unique solution. To see this, note that unique solutions can only exist if this equation can be rewritten in the standard form $\ddot{x}=F(x,\dot{x})$. This requires the invertibility of $\mathcal{G}(X,\cdot,X,\cdot)$, which gives the coefficients of~$\ddot x$ in~(\ref{Finslernull}), considered as a $4\times 4$ matrix. But since ${\cal G}(X,X,X,\cdot)=0$, this matrix does not have full rank. This is not surprising. Possible bifurcation points in the solutions should be expected, as we have seen that for wave covectors~$k$ that are double solutions of the Fresnel equation, the associated light ray~$X$ usually depends on the polarization. Spacetimes, or more generally any optical media with a simple (almost) metric constitutive relation are an exception to this rule: even though wave covectors~$k$ in this case are always double solutions of the Fresnel equation, the corresponding light rays~$X$ are unique. This is the deeper reason underlying the fact that the integral curves of light ray vectors become standard null geodesics in an almost metric background.

\subsection{Illustration: bimetric backgrounds}\label{subsec_birefringent}
Area metric spacetimes whose Fresnel tensor takes the bi-metric form
\begin{equation}\label{simplebirefr}
{\cal G}^{abcd}=g_{I}^{(ab}g_{II}^{cd)}\,
\end{equation}
constitute a special, but very important case of the general approach developed so far. In this section, we illustrate our general results on the local null structure and light propagation for this case, which is of direct relevance to our study of light deflection in spherically symmetric area metric backgrounds in section \ref{sec_lensingplanetary}. We will now prove the non-trivial fact that the metrics~$g_I$ and~$g_{II}$ also define the dual Fresnel tensor. The corresponding theorem further ensures that the correspondence between the directions of light ray vectors $X$ and wave covectors~$k$ in bi-metric backgrounds is everywhere bijective, in contrast to the existence of degenerate points in the generic case, see Theorem 2.

\vspace{6pt}\noindent\textbf{Theorem 3.} \emph{If the Fresnel tensor is of the bimetric form (\ref{simplebirefr}) in terms of two inverse metrics, then the dual Fresnel tensor is conformally related to the one constructed from the respective non-inverted metrics, i.e.,}
\begin{equation}\label{biref2}
{\cal G}_{abcd}\sim g_{I\,(ab}g_{II\,cd)}\,.
\end{equation}

\vspace{6pt}\noindent\emph{Proof.} In the first step we show that, if $g_{I}^{-1}(k,k)=0$ holds for a wave covector $k$, then the corresponding vector~$g_{I}^{-1}(k,\cdot)$ is a ray vector. For this we distinguish two cases. First, let $g_{I}^{-1}(k,k)=0$ but $g_{II}^{-1}(k,k)\neq 0$; then $\mathcal{G}(k,k,k,\cdot)=g_{II}^{-1}(k,k)g_I^{-1}(k,\cdot)$ is nonzero. By the argument used in the proof of Theorem 1, this shows that $g_I^{-1}(k,\cdot)$ is a ray vector. Second, consider the case $g_{I}^{-1}(k,k)=0$ and $g_{II}^{-1}(k,k)=0$. In this case one can show that $g_{I}^{-1}(k,\cdot)\sim g_{II}^{-1}(k,\cdot)$ and that $X=g_{I}^{-1}(k,\cdot)$ is a ray vector. The proof of this fact involves some algebraic detail and the explicit calculation of relevant components of the Fresnel tensor. Repeating the argument for the metric $g_{II}$ tells us that every wave covector yields a corresponding ray vector, null either with respect to $g_I$ or $g_{II}$. Hence $g_I(X,X)$ and $g_{II}(X,X)$, which are homogeneous polynomials of degree two, both divide the diagonal value $\mathcal G(X,X,X,X)$ of the dual Fresnel tensor. Since this polynomial is maximally of degree four, we obtain the decomposition $\mathcal G(X,X,X,X)=\alpha g_I(X,X)g_{II}(X,X)$ for some function~$\alpha$. Thus $\mathcal G(X,X,X,X)=0$ is equivalent to $g_{I\,(ab}g_{II\,cd)}X^aX^bX^cX^d=0$. By the Proposition in section \ref{sec_localnull} the claimed conformal equivalence follows.~$\square$

\vspace{6pt} Now consider the propagation of light on area metric backgrounds with bimetric null structure. Theorem~2 of the previous section then admits a simple reformulation, and we recover as special case a known result \cite{Perlickbook}: Finsler null geodesics reduce to null geodesics of either one of the two metrics~$g_{I}$ or~$g_{II}$. Indeed, starting from equation~(\ref{Finslernull}) and substituting the explicit expression $\mathcal G_{abcd}=g_{I (ab}g_{II\,cd)}$ for the dual Fresnel tensor, one obtains
\begin{eqnarray}\label{bigeodesic}
& g_{I}(X,X)\left[g_{II}(\nabla^{II}_X X,\cdot)+g_{II}(X,\cdot)\,X\!\left(\ln g_{I}(X,X)\right)\right] & \nonumber\\
+ & g_{II}(X,X)\left[g_{I}(\nabla^{I}_X X,\cdot)+g_{I}(X,\cdot)\,X\!\left(\ln g_{II}(X,X)\right)\right]  & =\,0\,,
\end{eqnarray}
where $\nabla^{I}$ and $\nabla^{II}$ are the Levi-Civita covariant derivatives associated to $g_{I}$ and $g_{II}$, respectively. The null condition ${\cal G}(X,X,X,X)=0$ implies that X must be null with respect to at least one of the two metrics, say $g_{I}(X,X)=0$. There are now two cases. If $g_{II}(X,X)\neq 0$, the equation above implies that $\nabla^{I}_X X=-X\!\left(\ln g_{II}(X,X)\right)X$, which shows that the integral curve $X=\frac{dx}{dt}$ is a (non-affinely parametrized) null geodesic of $g_{I}$. However, if also $g_{II}(X,X) = 0$, then the equation becomes trivial. These are exactly the points where the unique propagation along the Finsler geodesic breaks down.

On an area metric background with bimetric null structure there are generically  two different null vectors for a given spatial propagation direction of a light ray. We have seen that a specific polarization vector corresponds to each one of these null vectors~$X$, and is obtained as the solution of equation $G_{ijmn}X^iX^mQ^n=0$. Physically speaking, a bimetric background thus splits a generic electromagnetic wave into two polarized components that propagate along the geodesics of the two different metrics.

\section{Gravitational lensing}\label{sec_lensingplanetary}
In this section we study an application of our general results on light propagation on area metric backgrounds. We focus on the case of stationary spherically symmetric spacetimes, which is phenomenologically important in the solar system. After a discussion of the particular geometric structure of this class of spacetimes, we investigate observable physical effects such as birefringence and light deflection.

\subsection{Spherical symmetry}\label{sec_spherical}
The purpose of the present section is to present the generic form of a stationary, spherically symmetric area metric spacetime. We also calculate the form of the associated Fresnel tensor which is the relevant structure for light propagation. 

In order to implement symmetry conditions on an area metric, we recall from \cite{Punzi:2006nx} that an area metric isometry is a diffeomorphism $h: M \to M$ that preserves the area metric in the sense that for all smooth vector fields $U,V,A,B$ on $M$ and all $p\in M$, we have $G_{h(p)}(h_*U, h_*V, h_*A, h_*B) = G_p(U,V,A,B)$, where $h_*$ denotes the push-forward with respect to $h$. As in the metric case, it is useful to consider the generators of isometries. A vector field $X$ is a generator of a one-parameter family of isometries of an area metric manifold $(M,G)$ if and only if the Killing condition $\mathcal{L}_X G = 0$ holds. The Killing vectors of a given area metric manifold, together with the standard commutator, constitute a Lie algebra. We may now define spherical symmetry and stationarity for an area metric manifold, which is what we need to discuss area metric phenomenology at the solar system level. In analogy with the standard definition given in Lorentzian geometry, we call a four-dimensional area metric manifold spherically symmetric around a point $p$ if the isotropy group of $p$ is $SO(3)$, and the relative orbit of any other point $q\neq p$ around $p$ is topologically equivalent to a two-sphere. Further we call a spherically symmetric area metric manifold stationary if it admits a Killing vector field that commutes with the generators of $SO(3)$. Employing standard spherical coordinates $(t,r,\theta,\phi)$ these conditions amount to requiring that
\begin{eqnarray}\label{ssKilling}
& X_{0}=\partial_t\,,\qquad X_{1}=\sin\phi\,\partial_{\theta}+\cot\theta\cos\phi\,\partial_{\phi}\,,&\nonumber\\
& X_{2}=-\cos\phi\,\partial_{\theta}+\cot\theta\sin\phi\,\partial_{\phi}\,,\qquad X_{3}=\partial_{\phi} &
\end{eqnarray}
are Killing vectors of the area metric manifold. With some calculation one thus obtains the general form of a stationary spherically symmetric area metric spacetime. We display the area metric as a symmetric $6\times 6$ matrix $G_{MN}$, by considering Petrov indices $M,N$ that run over the antisymmetric index pairs $([tr],[t\theta],[t\phi],[r\theta],[r\phi],[\theta\phi])$:
\begin{equation}\label{sssareametric}
G_{MN}=
\!\left[\begin{array}{cccccc}
-A B \xi & 0 & 0 & 0 & 0 & (T+2S)r^2\sin\theta \\
  & -Ar^2 & 0 & 0 &(S-T)r^2\sin\theta & 0 \\
  &  & -Ar^2\sin^2\theta & (T-S)r^2\sin\theta & 0 & 0 \\
  &  &  & Br^2 & 0 & 0 \\
  &  &  &  & Br^2\sin^2\theta & 0 \\
  &  &  &  &  & r^4\sin^2\theta
\end{array}\right]\!\\,
\end{equation}
The empty slots are filled by symmetry, and $A$, $B$, $\xi$, $S$, $T$ all are functions of $r$. Note that this is just a convenient form to display all components of the \emph{four-dimensional} area metric. 

A general spherically symmetric area metric background is not metric-induced; this is only the case for $\xi=1$, $S=T=0$ when the inducing metric is the general stationary spherically symmetric metric $g_{ab}=\textrm{diag}(-A,B,r^2,r^2\sin^2\theta)_{ab}$. Another important remark concerns the signs of the area metric components. We require sections of constant $t$ to be spacelike, i.e. the restriction of the area metric to these three-dimensional sections must be positive-definite, and we require any area containing the vector~$\partial_t$ to be timelike, i.e., $G(\partial_t,\cdot,\partial_t,\cdot)$ should be negative definite if restricted to the three-dimensional complement of the span of $\partial_t$ in $TM$. It is indeed easy to show that the latter condition constitutes the appropriate generalization of the definition of a timelike vector in Lorentzian geometry, to which it reduces in the metric-induced case. These two conditions together require $A>0$, $B>0$, $\xi>0$.

The easiest way to calculate the Fresnel tensor, and to understand how a stationary spherically symmetric area metric background locally affects light propagation is to rewrite the area metric in the coframe $\theta^{\hat{0}}=\sqrt{A}dt,\theta^{\hat{1}}=\sqrt{B}dr,\theta^{\hat{2}}=r d\theta,\theta^{\hat{3}}=r\sin\theta d\phi$ and to define $\tau=T/\sqrt{AB}$ and $\sigma=S/\sqrt{AB}$. In the metric-induced case, this is the frame in which the inducing metric takes the Minkowski form. The dual Fresnel tensor is now easily calculated. One finds
\begin{equation}\label{birefr}
{\cal G}_{\hat{a}\hat{b}\hat{c}\hat{d}}=C(r)g^{+}_{(\hat{a}\hat{b}}g^{-}_{\hat{c}\hat{d})}\,,
\end{equation}
for the function $C(r)=\left[(1+\sigma^2)^2(\xi+4\sigma^2)\right]^{1/3}$, and the two metrics
\begin{equation}\label{gplusminus}
g^\pm_{ab}=\textrm{diag}(-\zeta^\pm,\zeta^\pm,1,1)_{ab}\,,\quad
\zeta^{\pm}=\frac{1}{2}\big(1+\xi+9\sigma^2\pm\sqrt{(1+\xi+9\sigma^2)^2-4\xi}\big)\,.
\end{equation}
It is easy to prove that because of $\xi>0$ the two quantities $\zeta^{\pm}$ are always real and positive. This means that the spherically symmetric area metric is actually bimetric, see section~\ref{subsec_birefringent}, because the two metrics~$g^\pm$ are Lorentzian. Moreover, $\zeta^{+}=\zeta^{-}$ if and only if $\sigma=0$ and $\xi=1$, i.e., when the area metric is of the almost-metric form. Therefore, the phenomenon of birefringence in spherically symmetric area metric spacetimes is equivalent to a deviation from almost-metricity.

In the remainder of this section, we are interested in observable effects in  the solar system which can be considered as a weakly gravitating system. Independent of the area metric gravity theory at hand the maximally symmetric vacuum is almost metric, and determined by the Minkowski metric and constant $\phi=\phi_0$. Keeping only the lowest order terms for every metric function, the expansion around the vacuum is
\begin{eqnarray}\label{approx}
A&\simeq& 1+\delta A\,,\qquad B\simeq 1+\delta
B\,,\qquad\xi=\zeta^+\zeta^-\simeq 1+\delta
\xi\,,\nonumber\\
\sigma&\simeq&\delta\sigma\,,\qquad \tau\simeq \phi_0+\delta\tau\,,\qquad \zeta^\pm\simeq 1+\delta\zeta^\pm\,.  
\end{eqnarray}

\subsection{Local effects of birefringence}
We now present a first analysis of the effects of deviations from metricity in spherically symmetric area metric spacetimes. In this section we study local effects  on the propagation of light rays, which however are hard to measure. A more detailed study of global effects on light deflection, which are more easily accessible to experimental tests, will be presented in the following section.

From the structure of the two metrics $g^{\pm}$ appearing in the Fresnel tensor (\ref{birefr}) it is evident that the null condition $\mathcal{G}(X,X,X,X)=0$ for radial light rays, which are of the form $X=X^{\hat 0}e_{\hat 0}+X^{\hat 1}e_{\hat 1}$, is the same for each metric, so that light travelling radially is not affected by birefringence, as one could expect by a simple symmetry argument. On the other hand, light rays propagating in non-radial directions can be null with respect to only one of the two metrics $g^\pm$. Without loss of generality, consider a future pointing ray vector $X^\pm$ in the $\theta=\pi/2$ plane. It must be of the form
\begin{equation}
X^{\pm}=\Big(\sqrt{(X^{\hat{1}})^2+(X^{\hat{3}})^2/\zeta^{\pm}},X^{\hat{1}},0,X^{\hat{3}}\Big).
\end{equation}
To each of the null vectors $X^\pm$ corresponds a polarization vector $Q^\pm$ that can be deduced from the relation $G_{abcd}X^b X^c Q^d=0$, up to a change of gauge $Q\mapsto Q+\lambda X$ which leaves the electromagnetic fields invariant. As an example consider the purely tangential case $X^{\hat{1}}=0$, where this relation simply reads
\begin{equation}
\sqrt{\zeta^{\pm}}Q^{\pm\,\hat{0}}-Q^{\pm\,\hat{3}}=0\,,\qquad 3\sigma
\sqrt{\zeta^{\pm}}Q^{\pm\,\hat{1}}+(1-\zeta^{\pm})Q^{\pm\,\hat{2}}=0\,.
\end{equation}
The first equation tells us that we can arrange for $Q^{\hat{0}}=Q^{\hat{3}}=0$ by a gauge transformation ${Q\mapsto Q+\lambda X}$. The second relation then fixes the two different polarization directions for $Q^\pm$. A general electromagnetic wave $F^{\#\,\hat a\hat b}=G^{\hat a\hat b\hat c\hat d}F_{\hat c\hat d}/2$ with given spatial direction of propagation will always be decomposable as $F^{\#}=\alpha X^+\wedge Q^+ + \beta X^-\wedge Q^-$; as discussed in section \ref{subsec_birefringent}, the two components propagate along the geodesics of $g^+$ or $g^-$, respectively.

For a given spatial direction of propagation, in terms of $X^{\hat 1}$ and $X^{\hat 3}$, two different coordinate velocities  are obtained, related to the polarization of the wave. In principle, this effect can be used to perform a laboratory test of area metric gravity, by simply measuring the coordinate velocities of light rays of different polarization. The effect is most prominent for light travelling in the tangential direction. Then $X^{\hat 1}=0$, and $c^\pm=X^{\pm\,\hat 3}/X^{\pm\,\hat 0}=\sqrt{\zeta^\pm}$. Writing $\Delta c=c^+-c^-$ and $c=(c^++c^-)/2$, and expanding around the area metric vacuum as in (\ref{approx}), we obtain
\begin{equation}
 \frac{\Delta c}{c}=\frac{\delta\zeta^{+}-\delta\zeta^{-}}{2}\,.
\end{equation}
Note that this quantity agrees, in our approximation, with the one measured by a laboratory observer at rest in the solar system \cite{Gabriel:1990qe}. This effect clearly presents a violation of the Einstein equivalence principle, in the form in which it states that gravitational effects can be cancelled locally by an appropriate choice of the local frame. Experimental bounds on~$\Delta c/c$ thus directly translate into a consistency requirement on spherically symmetric solutions of a theory of area metric gravity.

\subsection{Light deflection}
As we saw in the previous section, purely local effects of the non-metricity of an area metric background mainly manifest themselves as a dependence of the velocity of light on the propagation direction and polarization. These effects are much harder to detect than global ones, arising from the accumulation of small effects along the path of a light ray. In this section we investigate light deflection as an important example of such a global effect, and discuss the consequences for the viability of area metric gravity at the solar system level, by using the fundamental theorems on null geodesics derived in section \ref{sec_light}.

Recall that, according to theorems 2 and 3 proven in sections \ref{sec_global} and \ref{subsec_birefringent}, light rays in birefringent area metric backgrounds with Fresnel tensor (\ref{birefr}) follow null geodesics of either metric $g^+$ or $g^-$, compare equation (\ref{bigeodesic}). Starting from expression (\ref{gplusminus}) for these two metrics, we derive their form in the usual coordinate frame $(t,r,\theta,\phi)$:
\begin{equation}\label{coordmetric}
g^\pm_{ab}=\textrm{diag}(-A\zeta^{\pm},B\zeta^{\pm},r^2,r^2\sin^2\theta)_{ab}\,.
\end{equation}
Since area metric dynamics as discussed in
\cite{Punzi:2006nx} admits as maximally symmetric vacuum only a
Minkowskian, almost-metric background, we can assume the area
metric, and consequently the two metrics (\ref{coordmetric}), to
be asymptotically flat. Now standard machinery may be used to calculate the deflection angle of light by the spherically symmetric gravitational field~\cite{Weinbergbook}. As usual, the null geodesic light trajectories of a metric $g$ can be deduced as the Euler-Lagrange equations from the Lagrangian ${L=\frac{1}{2}g_{ab}\dot{x}^a\dot{x}^b}$, imposing the null condition $g_{ab}\dot{x}^a\dot{x}^b=0$, where the dot denotes differentiation with respect to an affine parameter along the curve. We assume the general form of a stationary spherically symmetric metric $g_{ab}=\textrm{diag}(-F(r),G(r),r^2,r^2\sin^2\theta)_{ab}$, and restrict (without loss of generality) to the case $\theta=\pi/2$. The null geodesics of $g$ (below we will replace the functions $F$ and $G$ by the expressions needed for $g^\pm$) satisfy
\begin{equation}
 \dot{t}=-\frac{E}{F(r)},\qquad \dot{\phi}=\frac{L}{r^2},\qquad
 \dot{r}=\pm\sqrt{\frac{1}{G(r)}\left(\frac{E^2}{F(r)}-\frac{L^2}{r^2}\right)}\,,
\end{equation}
where $E$ and $L$ are integration constants. Eliminating the affine parameter from the previous expressions for $\dot r$ and $\dot\phi$, we may solve for $d\phi/dr$. The total gravitational deflection angle of a light ray emitted and received at spatial infinity is now obtained by integration as
$\alpha(r_0)=2\int_{r_0}^{\infty} |d\phi/d
r|dr-\pi$, where $r_0$ is the radius of closest approach to the gravitating source. Defining the impact parameter~$b$ as $b=|L/E|$ we have $b^2=r_0^2/F(r_0)$. 

Since we are interested in solar system experiments, we can now assume slight deviations from a Minkowskian almost metric background, so that $F(r)\simeq 1+\delta F(r)$ and $G(r)\simeq 1+\delta G(r)$. Performing an expansion to first order, and using $x=r_0/r$, we find
\begin{equation}\label{alpha}
 \alpha(r_0)=\int_0^1 \frac{dx}{\sqrt{1-x^2}}\left[\delta G(x)+\frac{\delta F(x)-\delta
 F(1)}{1-x^2}\right].
\end{equation}
It is clear that a birefringent area metric background with two metrics $g^\pm$ will produce two different deflection angles for light rays of different polarization. In the example of a distant star, which can be considered as an unpolarized pointlike source, light deflection will hence produce a pair of differently polarized stellar images with a certain angular separation. This effect can be directly calculated from (\ref{alpha}), since the functions $F$ and $G$ are different for the two optical metrics $g^{+}$~and~$g^{-}$; for their deviations from the vacuum value, according to~(\ref{approx}), we have $\delta F^{\pm}=\delta A+\delta \zeta^{\pm}$ and $\delta G^{\pm}=\delta B+\delta \zeta^{\pm}$. The relevant quantity for the angular separation between the two polarized images then is
\begin{equation}
 \Delta\alpha(r_0)=\int_0^1\frac{dx}{\sqrt{1-x^2}}\left[\Delta \delta \zeta(x)+\frac{\Delta \delta \zeta(x)-\Delta
 \delta \zeta(1)}{1-x^2}\right]\!,
\end{equation}
where $\Delta \delta\zeta=\delta\zeta^{+}-\delta\zeta^{-}$. In terms of the first order expansion of the area metric functions we find $\Delta\delta\zeta\simeq\sqrt{\delta\xi^2+36\delta\sigma^2}$. Therefore this effect is first order in the deviation of the non-metric degrees of freedom from their background value.

To provide a specific example, it is reasonable to assume a power law radial dependence for the leading order term $\Delta \delta\zeta$, i.e., $\Delta\delta\zeta=(l_{\zeta}/r)^{\gamma}$ with $\gamma>0$. The quantity $l_{\zeta}$ would appear as a non-metric charge of the source, related to its energy momentum tensor through the gravitational field equations. With this assumption, we find
\begin{equation}
 \Delta\alpha(r_0)=\sqrt{\pi}\left(\frac{l_{\zeta}}{r_0}\right)^{\gamma}\frac{\gamma-1}{\gamma}
 \frac{\Gamma\left(\frac{1+\gamma}{2}\right)}{\Gamma\left({\frac{\gamma}{2}}\right)}\,,
\end{equation}
where $\Gamma(x)$ is the Euler gamma function. In the special case $\gamma=1$, the effects of birefringence cancel at leading order. For any other $\gamma$, the maximal attainable value of $\alpha(r_0)$ is obtained for minimal $r_0=r_\odot$, i.e., for light rays grazing the solar surface. Birefringence is not observed in experiment; therefore a consistent spherically symmetric area metric background must be such that the two differently polarized images of a given source cannot be resolved. Thus experiment gives an upper bound on the value of $l_{\zeta}$, which can be compared with the predictions from solutions of any given theory of area metric gravity. Assuming a best angular resolution of $0.001$ arc seconds for infrared interferometry (with the Very Large Telescope Interferometer VLTI, see \cite{VLTI}), we are led to bounds on $l_{\zeta}/r_{\odot}$, as summarized in the following table for various values of $\gamma$.
\begin{equation}\nonumber
\begin{array}{c|ccccccc}
\gamma & 1 & 2 & 3 & 4\\ \hline l_{\zeta}{}^{max}/r_{\odot} & \textrm{---} &
0.79\, 10^{-4} & 0.15\, 10^{-3} & 0.72\,10^{-2}
\end{array}
\end{equation}
Even better bounds are obtained from the study of depolarization effects. However, this requires statistical analysis and numerical modelling beyond the scope of this paper; compare, for instance, reference \cite{Solanki:2004az} where the case $\gamma=4$ leads to a bound of the order of magnitude $10^{-4}$.

\section{Conclusions}
In this article, we derived the equation governing light paths on area metric manifolds, independent of any of the various mechanisms \cite {Drummond:1979pp,Krasnov:2007uu,Krasnov:2007ky,Krasnov:2007ei,Schuller:2005ru,Maran:2005iu,Tamm,Obukhov:2002xa,Hehlbook,Hehl:2002hr,Rubilar:2007qm,Lammerzahl:2004ww,Itin:2004za,Hehl:2004zk,Kaiser:2004mz,Punzi:2008dv} that give rise to such backgrounds as refinements of metric geometry. Although an area metric presents a tensorial structure, light rays perceive those backgrounds as a Finslerian geometry. In contrast to Riemannian geometry, Finsler geometry provides a norm on each tangent space, rather than an inner product. Based on new results concerning area metric geometry, we calculated the Finsler norm seen by light rays in terms of the area metric, and found that light paths are curves which are null and stationary with respect to this Finsler norm, i.e., Finsler null geodesics.  

We learn most of what we infer about the large-scale structure of the universe from the propagation of light. So the result that light propagates along Finsler null geodesics on area metric backgrounds plays an important role in confronting any theory that effectively gives rise to such refined non-metric geometries with experimental data. Our analysis of light deflection illustrates the derivation of experimental bounds, which must be obeyed independent of the stipulated origin of the refined spacetime structure. 

\acknowledgments
The authors wish to thank Robert Oeckl and Valerio Bozza for helpful discussions and suggestions. RP acknowledges full financial support from the German Research Foundation DFG through grant WO 1447/1-1.  MNRW acknowledges full financial support through the Emmy Noether Fellowship grant WO 1447/1-1 from the German Research Foundation DFG. 


\end{document}